\documentclass[a4paper,twocolumn,11pt
,unpublished
]{quantumarticle}
\pdfoutput=1
\usepackage[utf8]{inputenc}
\usepackage[english]{babel}
\usepackage[T1]{fontenc}
\usepackage{amsmath}
\usepackage{amssymb}
\usepackage[numbers]{natbib}
\usepackage{dcolumn}
\usepackage{bm}
\usepackage{braket}
\usepackage{tabularx}
\usepackage[table]{xcolor}
\usepackage{threeparttable}
\usepackage{graphicx}
\usepackage{mathtools}
\usepackage{multirow}
\usepackage[colorlinks=true,linkcolor=blue,urlcolor=black,citecolor=blue,bookmarksopen=true]{hyperref}
\usepackage[caption=false]{subfig}
\usepackage{physics}
\usepackage{mathrsfs}
\usepackage{xifthen}
\usepackage{array}
\usepackage{enumitem}

\newcommand{\pnl}{Physical and Computational Sciences, Pacific Northwest National Laboratory, Richland, WA, 99354, USA}

\begin{document}

\title{AQM: A Refresh of the Abstract Qubit Model for Quantum Computing Co-design}
\author{Chenxu Liu}
\email{Email: chenxu.liu@pnnl.gov}
\affiliation{\pnl}
\orcid{0000-0003-2616-3126}

\author{Samuel A. Stein}
\affiliation{\pnl}
\orcid{0000-0002-2655-8251}

\author{Muqing Zheng}
\affiliation{\pnl}
\orcid{0000-0002-6659-9672}

\author{James Ang}
\affiliation{\pnl}
\orcid{0000-0002-7373-1889}

\author{Ang Li}
\email{Email: ang.li@pnnl.gov}
\affiliation{\pnl}
\orcid{0000-0003-3734-9137}


\maketitle

\begin{abstract}
Qubits are the fundamental building blocks of quantum information science and applications, whose concept is widely utilized in both quantum physics and quantum computation. While the significance of qubits and their implementation in physical devices have been extensively examined, now is the right time to revisit this understanding. In this paper, we introduce an abstract qubit model (AQM), offering a mathematical framework for higher-level algorithms and applications, and setting forth criteria for lower-level physical devices to enable quantum computation. We first provide a comprehensive definition of ``qubits'', regarded as the foundational principle for quantum computing algorithms (bottom-up support), and examine their requisites for devices (top-down demand). We then investigate the feasibility of relaxing specific requirements, thereby broadening device support while considering techniques that tradeoff extra costs to counterbalance this relaxation. Lastly, we delve into the quantum applications that only require partial support of ``qubits'', and discuss the physical systems with limited support of the AQM but remain valuable in quantum applications. AQM may serve as an intermediate interface between quantum algorithms and devices, facilitating quantum algorithm-device co-design. 
\end{abstract}

\section{Introduction} \label{sec:intro}

Quantum computing (QC) holds the potential for significant acceleration of classically challenging problems~\cite{Grover1996, Shor1994, Shor1997, Simon1997}, while quantum information processing (QIP) facilitates secure communication underpinned by quantum mechanics~\cite{BB84, Xu2020, Portmann2022}. Along with the rapid advancement of quantum technology is the growing selection of physical systems promising for QC and QIP. Despite these devices being rooted in diverse materials and fabrication processes, certain essential functionalities exist, such as initialization, control, and readout of quantum information, that every device is deemed to offer. These functionalities represent the commonality, or the criteria, necessary for a device to be capable of engaging in QC and QIP.

In order to examine whether a physical system can be a promising candidate for QC and QIP, DiVincenzo in his seminal work~\cite{DiVincenzo2000} proposed the well-known criteria. One of the key concepts, which will also be the main focus of our paper, is the concept of qubits. 

As the fundamental building blocks of gate-based quantum computers, the abstraction of qubits is necessary to build a model for quantum computers and support the upper-stack design. As described by Nielsen and Chuang, a qubit is defined as {\it a mathematical object, each of which has a state that is a unit vector in a two-dimensional complex vector space}~\cite{nielsen_chuang_2010}. Looking from the bottom of the QC design stack, this criteria implies also the requirements placed on the physical platforms. The physical platform should support such a system, abstracted as a qubit. To better understand such a qubit in realistic physical systems, especially when the physical degrees of freedom are more than a qubit required, in Ref.~\cite{Viola2001}, Viola {\it et al.} proposed that: in addition to the operator algebra, the physical system should also support (1) universal control, (2) initialization, and (3) read-out capabilities to build an operational qubit.

It has been more than 20 years since DiVincenzo's qubit criteria. During the recent development of QC and QIP, new opportunities and challenges have arisen. Firstly, although a large integration of physical qubits has been demonstrated~\cite{IBMQ2025_Roadmap, Bluvstein2024, IBMQ_condor}, a perfect logical qubit has not been built yet. We still need to take advantage of hardware features to maximally improve the QC development. Second, some recent quantum algorithms can still function without the full support of the AQM, and hence the rigorous AQM has the potential to be further relaxed. Beyond the commonly used quantum circuit model, various universal computation models exist, such as measurement-based quantum computation (MBQC)~\cite{Raussendorf2001, Raussendorf2003, Browne2005, Briegel2009} and quantum random walk~\cite{Kadian2021, Andraca2012, Childs2009, Singh2021}. These models have unique requirements for physical systems. Lastly, with the recent progress of quantum co-design~\cite{Shi2020, Tomesh2021}, partially supporting AQM can provide benefits to both quantum algorithm design and hardware development. How to leverage the AQM for algorithm-device co-design is still unclear. These emerging topics call for a reconsideration of the AQM.

In this paper, we address these challenges by revisiting the AQM. We begin by presenting a comprehensive definition of qubits, essential for quantum algorithms (bottom-up support), and analyze their device requirements (top-down demand). We explore the opportunities of relaxing specific requirements to expand device support, considering techniques that trade off extra costs to compensate for the relaxation. Finally, we examine quantum applications that only need partial AQM functionalities and discuss physical systems with limited AQM support yet remain valuable in quantum applications.

This paper is organized as follows. In Sec.~\ref{sec:qubit_model}, we define the abstracted qubit model. In Sec.~\ref{sec:release_physical}, we discuss how the physical requirements can be relaxed and compensated. In Sec.~\ref{sec:beyond}, we focus on relaxing the complete AQM from both the upper and lower stacks opportunities. Finally, we conclude in Sec.~\ref{sec:summary}.

\begin{figure}[htbp]
    \centering
    \includegraphics[width=0.45 \columnwidth]{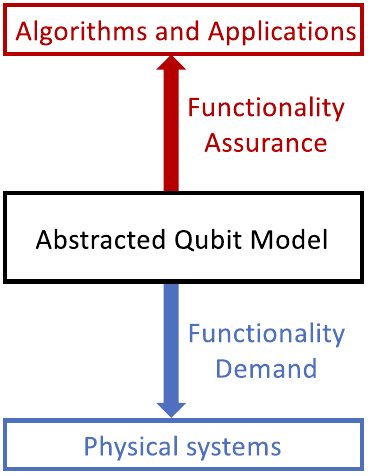}
    \caption{The abstracted qubit model in the design stacks of quantum information processing. The abstracted qubit model contains the mathematical description of qubits, which can be used in the upper stacks, while it places physical requirements that need to be fulfilled by the lower stacks. The complete AQM can assure the function execution in quantum algorithms (the red arrow), while it also places demands on the physical systems that construct qubits (the blue arrow).} 
    \label{fig:AQM_stacks}
\end{figure}

\section{Abstract model of qubits} \label{sec:qubit_model}

We present the definition of qubits and construct the AQM. The AQM design stack is shown in Fig.~\ref{fig:AQM_stacks}. Specifically, we consider how qubits can support the upper stacks of QC and QIP systems by defining the mathematical description of the states and the operations on qubits in Sec.~\ref{subsec:math}. Then, we discuss the requirements of AQM placed on the physical systems in Sec.~\ref{subsec:physical}. 

\subsection{Mathematical description of qubits} \label{subsec:math}

We first extract the essential properties of qubits. We point out how a qubit should be viewed in quantum algorithms and applications. 
\begin{itemize}[noitemsep]

\item {\bf State of qubits:} The state of a qubit can be represented as a complex two-dimensional vector with the unit norm, same as Ref.~\cite{nielsen_chuang_2010}. The state of $n$ qubits is a complex vector with dimension $2^n$. Assuming the state of qubit A is $\ket{A}$, while the state of qubit B is $\ket{B}$, the state of both qubits is $\ket{A} \otimes \ket{B}$, where $\otimes$ represent the Kronecker product of two vectors. 

\item {\bf Operations on qubits:} The quantum manipulation on qubits can be represented as a unitary matrix $U$, i.e., $U \, U^\dagger  = U^\dagger \, U = I$, where $I$ is an identity matrix. The dimension of the unitary matrix is $2^n \times 2^n$ if the operation is acting on $n$ qubits. Any unitary matrices ($U$) with the given dimension should be obtainable by qubit operations ($\tilde{U}$) with a global phase factor, $U = e^{i \phi} \tilde{U}$.

\item {\bf Measurements on qubits:} A measurement on $n$ qubits is represented by a set of $2^n\times 2^n$ matrices, noted as $\mathbf{M} = \{ m_1, m_2, ... \}$, where the matrices satisfies $\sum _{\forall m \in \mathbf{M}} m^\dagger m = I$. After performing the measurement, the state of the qubits is transformed into $\ket{\psi_j} = m_j \ket{\psi_0} / \sqrt{p_j}$, where $p_j = \bra{\psi_0} m_j^\dagger m_j \ket{\psi_0}$ is the probability of getting $j$-th measurement outcome ($j=0, \, 1$ for a projective measurement on a single qubit). If the measurement outcome is not observed, the state of the qubit becomes a classical probabilistic superposition of all the possible states $m_j \ket{\psi_0}$.
\end{itemize}

{\bf Qubit vs Qudit:} With the possibility of data compression and the advantages of simulating the dynamics with other symmetry groups, qudits become an attractive generalization and alternative to qubits~\cite{Wang2020, Chi2022, Liu2022Qudit, Cuadra2022}. A qudit is a system with $d$ states, where $d$ can be more than $2$. The mathematical definition discussed above can be easily generalized to qudit systems by redefining the dimensionality of a single qudit to $d$, while the operation matrix dimensions are $d \times d$. In the following of our discussion, we mainly focus on qubits, while the generalization to qudit systems is a straightforward simple linear mapping.

\subsection{Physical requirements for constructing qubits} \label{subsec:physical}

Recall DiVincenzo's five criteria~\cite{DiVincenzo2000} for examining physical systems for QC:
\begin{enumerate}[noitemsep]
    \item A scalable physical system with well-characterized qubits.
    \item The ability to initialize the state of the qubits to a simple fiducial state.
    \item Long relevant decoherence times, much longer than the gate operation time.
    \item A ``universal'' set of quantum gates.
    \item A qubit-specific measurement capability.
\end{enumerate}
and two extra criteria for quantum communication,
\begin{itemize}[noitemsep]
    \item[6.] The ability to interconvert stationary and flying qubits.
    \item[7.] The ability to faithfully transmit flying qubits between specified locations.
\end{itemize}

Subsequent works refine qubit requirements and explore building qubits for QC and QIP~\cite{Viola2001, DuckeringThesis}. Here we focus on DiVincenzo's criteria because of the wide adoption in QC. The ideal qubit model contains:
\begin{itemize}[noitemsep]
    \item {\bf Rule 1: States.} A qubit should have {\it $2$ quantum states} which can be addressed at any time of interest.
    
    \item {\bf Rule 2: Operations.} The capability of performing operations on qubits requires the physical system that makes these $n$ qubits to be {\it completely controllable}~\cite{Dong2010QControl, Alessandro2021introduction}. Obviously, with the complete controllability of the physical systems, universal quantum computing can be supported.

    \item {\bf Rule 3: Connectivity.} The capability of performing any unitary operations on qubits requires the qubits to have all-to-all connectivity.

    \item {\bf Rule 4: Coherence.} The qubit should have {\it infinite long coherent time and experiences no other incoherent errors}, which ensures that the state of the qubits at the time of interest can always be represented as a $2$-dimensional complex vector with unit length.
    
    \item {\bf Rule 5: Readout.} In addition, as projective measurements and post-selection results can be well described by the measurement formalism~\cite{nielsen_chuang_2010}, with the above measurement operation support, {\it the qubit states can be readout} into classical information. We also implicitly require that the measurement is error-free, meaning that the extracted classical information is accurate.
    The initialization of the qubits to a known state (noted as $\ket{\psi_0}$) can also be realized through this measurement operation by defining the measurement operator set $\mathbf{M} = \{\dyad{\psi_0}{j}\}$ for all basis states of the qubit systems $\ket{j}$. 
\end{itemize}

\section{Partial support of the AQM from physical systems} \label{sec:release_physical}

\begin{table*}[tbp]
\caption{\label{tab:requirement_releasing} Comparison of the ideal AQM and the partial AQM. We also consider the techniques to compensate for the imperfection and their corresponding costs.} 
\scriptsize
\begin{tabularx}{\textwidth}{p{3cm}|p{3cm}|p{3.5cm}|X}
\hline \hline
Ideal model & Partial model & Compensation techniques & Cost \\
\hline
Two quantum states & Two quantum states & \\
\hline
Infinitely long coherence time & Finite but relatively long coherence time & Quantum Error Detection and Correction & More physical qubits and more gate and measurement operations \\
& & Quantum error mitigation & Extra quantum algorithm implementations, and post-processing loads.\\
& & Dynamical decoupling & Extra control complexities \\
\hline
No quantum operation errors & Gate imperfection are allowed & Quantum Error Detection and Correction & More physical qubits, gates, and measurement operations \\
\hline
Directly support any unitary operations. & A discrete universal gate set is available. & Gate decomposition & Increased gate operations, which require more operation time. Needs more qubits if the gate set is computationally universal. \\
\hline
All-to-all connectivity. & All qubits can be connected. & Gate routing & More physical operations and extra times. \\
\hline
General measurements are accessible. & Projective measurements can be performed, even indirectly. & Measurements with auxiliary qubits & Extra qubits and quantum gates to construct a general measurement. \\
& & Routing for measurements &  Swapping the quantum states to the qubits requires extra gate operations and time.\\
\hline
\hline
\end{tabularx}
\end{table*}

\begin{figure}[ht]
    \centering
    \includegraphics[width = 0.45 \columnwidth]{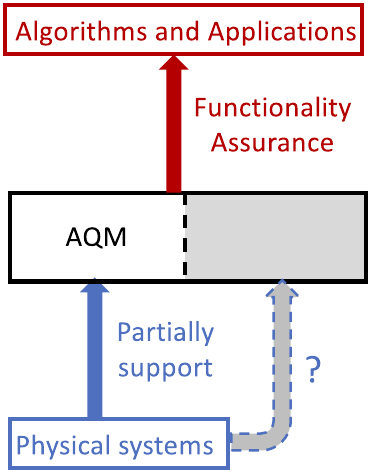}
    \caption{The partial support of the AQM from the physical systems. With compensation techniques, the physical systems can support the complete AQM with extra costs on time and resources. The physical systems can partially support the AQM (the blue arrow), while the AQM can assure the function implementation in quantum algorithms (the red arrow). Compensation techniques can be applied to support the complete AQM (the gray arrow).}
    \label{fig:compensation}
\end{figure}

The complete AQM is demanding. Most present physical systems can only support the AQM partially, as shown in Fig.~\ref{fig:compensation}. Nevertheless, full support of a computation qubit in the AQM is still possible by using the relaxation with compensation techniques. However, these techniques use extra physical resources, time, and complexity cost. These compensation techniques answer the question mark shown in Fig.~\ref{fig:compensation}. In the following, we break down the requirements to construct a qubit from physical systems and present the available techniques to compensate for the ``imperfection'' of the physical platforms. 

\paragraph{{\bf a. Relaxing the qubit state requirement:} The qubit should have two quantum states.} The existence of two quantum states for a qubit is necessary. One commonly seen example is in trapped ion or neutral atom systems, where the quantum levels of the ions and atoms are used to encode quantum information. Another example is to use the presence or absence of a particle (photons or electrons, etc.) in a single mode as the two quantum states of the qubits. We will discuss this in more detail in Sec.~\ref{subsec:photon}. There may be more than two quantum states in a certain physical system~\cite{Viola2001}. Other than simply picking two distinct states from all available quantum states, the quantum states need to be controllable (see the following discussion about Operation). 

However, some physical platforms may not have sufficient quantum states that fulfill even the minimum requirements of a single qubit.  The states spanning the state space should be addressable and controllable. When a single physical system is not sufficient to provide such a controllable state space, however, the composite system of multiple physical systems can be utilized to form a qubit. The composite system can consist of multiple copies of the same physical systems, e.g., the presence or absence of a single electron in a quantum dot cannot be encoded as qubit states due to the lack of coherent manipulation techniques in the Hilbert space spanned by these two states~\cite{DiVincenzo2000}. In contrast, a different physical system can be utilized to provide the necessary qubit operations. e.g., the presence or absence of a microwave photon can be encoded as a qubit with the help of a superconducting qubit~\cite{Mariantoni2011, Mariantoni2011Science, Besse2020, Reuer2022, Gao2019}.

We note that this requirement is weaker than the requirements of single-qubit completely controllable (Rule 2)~\cite{Dong2010QControl, Alessandro2021introduction}. Despite incomplete control over the physical system, if a set of computational universal gates can be executed, and the qubit state can be manipulated to a specific state (e.g., $\ket{0}$), the construction of a qubit is possible. The arbitrary state of the qubit Hilbert space can be approximately synthesized with the extra cost of quantum gate operations (see the discussion of Operations).

The quantum information encoded in a qubits Hilbert space can be transmitted via various physical systems. In measurement-based quantum computing (MBQC)~\cite{Raussendorf2001, Raussendorf2003, Raussendorf2007, Briegel2009}, for instance, information is "teleported" to connected physical qubits in resource states, with the physical qubits being destroyed during computation, especially in photonic systems. 

\paragraph{{\bf b. Relaxing the operation requirement:} The qubit should support a set of computational universal gates.} 
It is quite costly to require any unitary matrices on an arbitrary number of qubits to be directly implementable as different operations. What is the minimum requirement on the set of supported operations to allow us to at least approximate a given unitary matrix on a number of qubits? This question motivates the research surrounding universal gate sets for QC. It turns out that a discrete set of gates (unitary matrices) on a small number of qubits is enough ($n \leq 3$), e.g., the set of Hadamard gate, CNOT gate, phase gate, and $\pi/8$ gate, is universal~\cite{nielsen_chuang_2010}.

However, relaxing this requirement also results in extra costs. Gate decomposition which decomposes the required unitary matrix into the directly supported gate operations in the gate set can be rather costly. The gate decomposition requires an increasing number of quantum gates applied to the qubits. For example, the Solovay-Kitaev theorem showed that approximating an arbitrary unitary matrix in $SU(d)$ using a discrete set of gates that generate a dense subgroup of $SU(d)$ requires $O(log^c(1/\epsilon))$ gates, where $c$ is a constant and $\epsilon$ is the accuracy~\cite{nielsen_chuang_2010, dawson2005solovaykitaev}. The second aspect is that the different universal gate sets supported by the respective physical system can cost extra quantum gates and qubits for the same computation, as highlighted by the difference between `computational universal' and `strictly universal'~\cite{aharonov2003simple}. One example is the gate set that consists of the Toffoli gate and the Hadamard gate. This gate set is universal~\cite{aharonov2003simple, shi2002toffoli}, but a general unitary on a $n$-qubit system cannot be decomposed with arbitrary accuracy using just Toffoli and Hadamard gates. Instead, the computation process of that unitary operation can be simulated within an arbitrarily small error~\cite{aharonov2003simple}. However, it costs polynomially many more qubits and gates. Similar discussions about universality can be found in qudit systems~\cite{brylinski2001universal, Wang2020}. We stress that the cost of decomposing target unitaries on the qubit system should be manageable, e.g., not scaling exponentially more time and physical qubits.

Nevertheless, there are other QC models other than the quantum circuit models. For example, in the MBQC scheme~\cite{Raussendorf2001, Raussendorf2003, Raussendorf2007, Briegel2009}, universal computing is driven by the measurements on a prepared entangled state. In this model, the computation can still be mapped back to the circuit model, where the flow of quantum information can be treated as qubits that carry quantum information. The physical systems that support MBQC can still be treated as supporting full universal control, though indirectly with the help of measurements and pre-existing quantum entanglement. 

\paragraph{{\bf c. Relaxing the connectivity requirement:} The qubits need to be connected.} 
For a QC system supporting a universal gate set applicable up to $n$ qubits, a prerequisite for using these qubits in general QC algorithms is that any subset of $n$ qubits must be capable of performing gates from the universal set. 
Note that all-to-all connectivity is not necessary. For example, if the universal gate set contains CNOT, Hadamard, Phase, and $\pi/8$ gate~\cite{nielsen_chuang_2010}, this is equivalent to requiring that all qubits can perform the Hadamard, phase and $\pi/8$ gates, and any two qubits can find a path in their connection graph. 
Even if two qubits are not directly connected, a CNOT gate between them can be implemented using SWAP gates along the path, decomposable into three CNOTs. This less stringent connectivity requirement incurs additional costs compared to all-to-all connectivity.

In the new paradigm of distributed QC models, the ``connection'' between the two qubits we discussed above is not necessarily a direct coupling between two physical qubits. Instead, quantum communication between two physically remote qubits enables remote gate operations~\cite{Cuomo2020, Cacciapuoti2020}, which relies on a reliable channel to transmit flying qubits from one end to the other, or a reliable source of entangled resource states (like Bell states) to be used for gate teleportation~\cite{Gottesman_1999, Huang2004, Wan2019, Chou2018, nielsen_chuang_2010}. This should also comply with DiVincezo's criteria for quantum communication (criteria~6 and~7, see Sec.~\ref{subsec:physical}).

\paragraph{{\bf d. Relaxing the coherence requirements:} The qubit should have a finite but long enough coherence time.} 
Quantum coherence is critical for quantum computing. However, physical systems can hardly be completely isolated. The interaction with the external environment causes decoherence of the quantum systems. The coherence time gives a time scale for preserving the quantum coherence. These unwanted interactions also introduce decoherence errors to the quantum gates applied to the systems. Therefore, the coherence time should be long enough for the gate operations time to perform a meaningful QC on a physical system. This motivates the development of noisy intermediate-scale quantum (NISQ) algorithms, e.g., variational quantum algorithms~\cite{Cerezo2021}.

Another strategy is to suppress noise and extend the coherence time of the logical qubits by quantum error correction (QEC) codes~\cite{nielsen_chuang_2010, Calderbank1996}. By combining multiple physical qubits into a logical qubit, QEC codes use entangled states as qubit states to detect and correct errors. Increasing the code distance, achieved by using more physical qubits per logical qubit, allows for exponentially suppressing logical error rates as long as noise-induced error rates remain below the code threshold. What's more, QEC not only extends the coherence time of logical qubits but also suppresses noises from other sources, e.g., imperfect control during gate implementation and cross-talk between adjacent qubits, etc.

\paragraph{{\bf e. Relaxing the readout requirement:} The qubit should have a way of performing measurements on qubit states.} Compared to the mathematical description of the general measurements on ideal qubits, a quantum system to build a qubit may only support projective measurements along the computational basis. It is possible to construct general measurements using auxiliary qubits, entangling gates, projective measurements on the auxiliary qubits, and quantum gates conditioned on the projective measurement outcomes. 

Furthermore, even the direct projective measurements on the physical systems can be relaxed if we consider heterogeneous QC architectures, where the qubits used in different functional units inside the QC units~\cite{stein2023microarchitectures, stein2023multimode, liu2023quantum} can be different. The role of the qubit used in the quantum processing units and quantum memory units is to process and store quantum information, which means direct measurements on them are not necessary. We can think of using another physical system, that supports qubit measurement functionalities and can interact with the qubits in the quantum memory modules to extract the stored quantum states. Therefore, when the measurement of the quantum information is needed, it is performed on the second type of qubits by swapping the information out from the computing qubits. A more detailed discussion is in Sec.~\ref{subsec:qm_application}.

Table~\ref{tab:requirement_releasing} summarizes our discussion on releasing requirements for an ideal AQM. We also cover compensation techniques and their associated costs. If a physical system meets the requirements of a restricted AQM and supports the compensation techniques, it's suitable for constructing a large-scale general-purpose quantum computer. Such a system can serve as a candidate for building a qubit sufficient to support any quantum algorithms.

\begin{figure}[h]
    \centering
    \includegraphics[width = 0.45 \columnwidth]{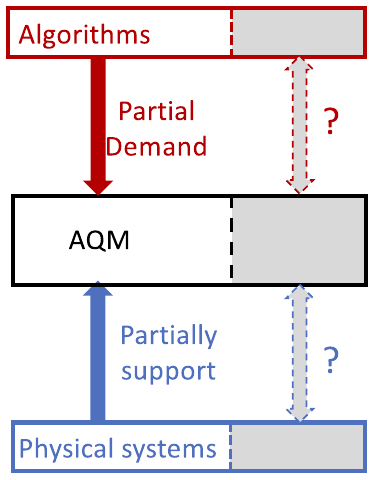}
    \caption{The algorithm-device co-design. Quantum algorithms and applications may only have partial demand on the AQM, while the physical systems can only have partial support of the AQM. The co-design of quantum algorithms and physical systems can be beneficial. While physical systems can only offer partial support for the AQM (the blue arrow), there are quantum algorithms that only partially require the AQM (the red arrow). The AQM provides the necessary tool for quantum co-design, which matches the demand from the quantum algorithms and the support from the physical systems (the gray arrows).}
    \label{fig:codesign}
\end{figure}

\section{AQM for Algorithm-Device codesign} \label{sec:beyond}

In this section, we discuss the algorithm-device co-design opportunities demonstrated by Fig.~\ref{fig:codesign}. Specifically, a quantum application may not require the complete support of the AQM, and hence a physical system that fulfills its requirements can still be useful, even without building a qubit, and vice versa. Therefore, the algorithm-device codesign opportunities by partially breaking the abstraction layer of qubits can be beneficial for the current stage of QC system design. Specifically, in Sec.~\ref{subsec:release_alg}, we discuss the algorithm and applications that can be supported without the complete AQM, while in Sec.~\ref{subsec:supporting_AQM}, we discuss different physical systems supporting the AQM or quantum applications. In Sec.~\ref{subsec:codesign}, we discuss co-design opportunities of the physical systems.

\begin{table*}[tbp]
\caption{\label{tab:algorithms} Comparison of the demand of quantum algorithms and applications on the AQM. We focus on quantum annealing (QA), quantum random walk (QRW), measurement-based quantum computing (MBQC), analogue quantum simulation (AQS), and quantum memory (QM). $\checkmark$ means the demand is similar to the AQM.} 
\small
\begin{tabularx}{\textwidth}{p{1.2 cm}|p{3 cm}|p{2.2 cm}|p{3.5 cm}|p{1.5 cm} |p {3.2 cm}}
\hline \hline
    & States    & Operation     & Connectivity  & Coherence     & Readout \\
\hline
QA  & $\checkmark$ & Relaxed & Critical to maintain high connectivity & $\checkmark$ & Computational basis\\
\hline
QRW & Encoding methods can be different. & Released & Based on the problem, can be relaxed. & $\checkmark$ & Computational basis \\
\hline
MBQC & Physical qubits are not always alive. & Relaxed (between physical qubits) & Resource state generation. & $\checkmark$ & Arbitrary basis \\
\hline
AQS  & $\checkmark$ & Relaxed & Based on the problem. & $\checkmark$ & Computational basis (depending on problems).\\
\hline
QM   & $\checkmark$ & Relaxed & Connectivity to computing component is necessary.  & Long coherence time. & Relaxed \\
\hline
\hline
\end{tabularx}
\end{table*}

\subsection{Applications without the complete AQM} \label{subsec:release_alg}

In this section, we consider three computation models, including quantum annealing, MBQC, and quantum random walk for computation. In addition, we also focus on the application of analog quantum simulation and quantum memory, which do not necessitate the complete AQM. The demand for these quantum applications is summarized in Table.~\ref{tab:algorithms}.

\subsubsection{Quantum Annealing} \label{subsec:annealing} 

Analogous to the classical annealing method, quantum annealing is a QC paradigm to solve combinatorial optimization problems~\cite{Das2008, Hauke2020, Albash2018}. Unlike the quantum circuit computing model, quantum annealing is an analog quantum computation framework. The optimization problem is encoded into the device Hamiltonian, whose ground state corresponds to the global optimum of the problem. Solving the problem corresponds to annealing the system states from an initial state to the ground state of the problem Hamiltonian. 

The device Hamiltonian can be described by
\begin{equation}
    H = \lambda_0 (t) H_0 + \lambda_1 (t) H_1,
\end{equation}
where $\lambda_{0,1}(t)$ are two tunable parameters, $H_{1}$ is the Hamiltonian that encodes the optimization problem, $H_{0}$ is another Hamiltonian which does not commute with the terms in $H_{1}$. In the beginning, $\lambda_0 = 1$ and $\lambda_1 = 0$, and the system is initialized into the ground state of $H_{0}$. During the annealing process, $\lambda_0$ is slowly decreased to $0$, while $\lambda_1$ is increased to $1$. As long as the change of the system Hamiltonian is slow enough, based on the adiabatic theorem, the state of the system will be tuned to the ground state of $H_1$, which solves the optimization problem.

Therefore, quantum annealing does not require the support for Rule 2 of the AQM. As the computation is not driven by quantum gate operation, there is no need to support a full universal gate set. Instead, as long as the physical device can realize the Hamiltonian $H_0$ and $H_1$ that encodes the optimization problem, e.g., the nearest-neighbor spin-spin coupling and single-qubit Pauli-X rotations~\cite{Johnson2011}, and can slowly tune the system Hamiltonian and readout the final state of the system, the system can support the quantum annealing algorithm to solve the specific type of optimization problems.

For example, one type of physical system commonly considered in quantum annealing is the Ising spin glass, where $H_0$ contains single-site $\sigma^{(x)}$ terms, while $H_1$ contains two-site $Z$ couplings, i.e., $\sigma^{(z)}_i \sigma^{(z)}_j$. Although the coupling terms are restricted to two-body interactions, NP-hard problems can be cast into quadratic forms using ancilla~\cite{Hauke2020, Lucas2014}. A survey of NP problems mapping to spin systems can be found in Ref.~\cite{Lucas2014}. Note that realizing the Hamiltonian terms does not lead to universal control of individual spin in the system. 

In order to keep the resource cost low when encoding the optimization problems into a quantum annealing setup, the physical systems need to have high connectivity, which is Rule 3 of the AQM. However, there are embedding techniques that allow mapping a quantum annealing problem onto physical systems with limited connectivity~\cite{djidjev2018efficient, Zaribafiyan2017, Lechner2015}.

In conclusion, the quantum annealing applications require \textbf{Rule 1 and 4} of the complete AQM, while the \textbf{Rule 2 and 5} can be relaxed. In addition, the \textbf{Rule 3} is also critical to keep the resource cost low.

\subsubsection{Quantum random walk.}

Quantum random walk is a computational paradigm for QC, especially for solving graph problems. There are broad categories of quantum walk computations. In this section, we do not aim to give a thorough review of quantum random walk. For a review of quantum random walk for QC, we refer to Refs.~\cite{Kadian2021, Andraca2012}. Instead, we focus on the coined discrete-time quantum walk as an example.

Compared to the classical random walk model, where the walker stochastically chooses the movement, in the coined discrete-time quantum walk, there are two quantum systems, one is used as a coin, while the other system is the walker system. The walker's movement is determined by the coin state. The walker's trajectory is entangled with the coin states and different trajectories are coherently superposed.

For instance, when a quantum walker working on a 1D line, who can only walk to the right or left. The coin system can be a two-level system, while the walker's state encodes the position of the walker on the line. The walk operation can be represented as a unitary acting on both systems,
\begin{equation}
    U = \dyad{0}^{c} \otimes U_0^w  + \dyad{1}^{c} \otimes U_1^w,
\end{equation}
where $U_0^w$ and $U_1^w$ are unitary operations acting on the walker system representing the movements. To encode different probability distributions of the walker, the coin system should be able to apply a general unitary in the Hilbert space (a general $SU(2)$ operation if it is a qubit). In addition, it is necessary to perform measurements on the coin and walker systems to extract the computation results.

The quantum walk model can be used to implement universal QC~\cite{Childs2009, Singh2021}. The quantum random walk model needs a different set of requirements compared to the AQM. For example, quantum information can be encoded not into the internal state of the qubit, but instead, into the spacial modes that can hold a physical particle~\cite{Tang2018}. Other than supporting a universal gate set on qubits, quantum walk computation only requires that the physical qubit can hop between different spatial modes that encode the quantum information to carry out the computation. 

In summary, quantum walk applications require \textbf{Rule 4} of the AQM. However, they have distinct requirements from the general AQM. Adaptable to problem demands, the quantum walk can operate on systems with more than two quantum states by encoding the coin system into a multi-level system based on walking choices, which differs from the \textbf{Rule 1} of the AQM. \textbf{Rule 2} of the AQM is relaxed, as universal controllability and a universal gate set are not essential while controlled-walk operation suffices. \textbf{Rule 3} can also be partially relaxed, where the unique connectivity involves linking the systems encoding coin and space degrees of freedom (DOFs). Full connectivity is not required for unreachable space DOFs. For readout, the quantum walk model primarily emphasizes distribution in space DOFs~\cite{Gong2021QW, Esposito2022}, making complete quantum information extraction (e.g., state tomography) unnecessary. Therefore, \textbf{Rule 5} is also relaxed.

\subsubsection{Measurement-based quantum computation.} \label{subsec:MBQC}
MBQC is a computation model different from the quantum circuit computation model~\cite{Raussendorf2001, Raussendorf2003, Browne2005, Briegel2009}. In MBQC, the computation is driven by measurements of qubits in an entangled state of qubits. This entangled state is the resource of MBQC. A two-dimensional and higher-dimensional cluster state can be used as the resource state of MBQC~\cite{Raussendorf2003, VanderNest2006, Brown2020}.

As demonstrated in Ref.~\cite{Raussendorf2003}, a set of universal gates can be equivalently performed in MBQC. Therefore, MBQC is equivalent to the quantum circuit computation model and it can support universal QC. However, compared to the requirements placed on the AQM, MBQC releases a few requirements. The physical qubits used in the MBQC are not accessible at all times, and they do not need to support a universal gate set for computation. Furthermore, the connectivity requirements are also different from the AQM, as it only requires that the physical qubits need to be able to prepare a specific resource state for MBQC. During the computation, connectivity between different physical qubits is not necessary.

MBQC requires fast and high-fidelity single-qubit measurements on the qubits, and the measurement basis can be easily adjusted. Additionally, although on-demand entangling operations aren't essential during computation, preparing an entangled state of all qubits can be demanding. Moreover, due to MBQC's nature, performing the same algorithm requires a larger number of physical qubits compared to the circuit computation model.

In summary, MBQC requires \textbf{Rule 4 and 5} of the AOM. In contrast, MBQC relaxes \textbf{Rule 1} of the AQM, where the physical qubits do not need to be alive throughout the computation. \textbf{Rule 2} of the AQM is also relaxed, as the physical qubits do not need to perform entangling gates. \textbf{Rule 3} is supported in a different manner, which refers to the connectivity of the resource state. 

\subsubsection{Analogue Quantum Simulation}

Quantum simulation utilizes the power of QC systems to model the behavior of a desired target system. By measuring the quantum system, we can extract physical properties of the target systems that are difficult to compute using classical methods.

There are two main quantum simulation categories based on the type of quantum systems used. One involves digital quantum computers equipped with algorithms to compute physical properties, such as the VQE algorithm designed for determining molecular ground states and energies~\cite{Cerezo2021}. The other method focuses on employing quantum devices for analogue simulations~\cite{Monroe2021, Lv2018, Ebadi2021, Daley2022}, which is the emphasis of this subsection.

Analogue quantum simulations can address low-energy steady states or dynamics of physical systems. Solving the ground state of a physical system can use adiabatic evolution, akin to the discussion of the quantum annealing method (see Sec.~\ref{subsec:annealing}). Alternatively, when the problem directly maps to a quantum system, i.e., the quantum simulator can realize the target Hamiltonian, manipulating the simulator can simulate the dynamics of the target system. Therefore, the measurement of the quantum simulator can help to reveal the physical properties of the target problem. For example, the Ising model can naturally map to a neutral atom lattice~\cite{Ebadi2021}. The quantum many-body phase transition can be directly observed by manipulating the atom interactions~\cite{Ebadi2021, Bloch2008}.

In summary, analogue quantum simulation still demands \textbf{Rule 1 and 4} of the AQM. As the simulator does not need to support a universal gate set for computing, or achieving universal control on each qubit, as long as the target Hamiltonian can be realized, \textbf{Rule 2} of the AQM is relaxed. As the connectivity and the readout capability depend on the physical problems being simulated, \textbf{Rule 3 and 5} of the AQM have the potential to be relaxed.

\subsubsection{Quantum Memory} \label{subsec:qm_application}

Quantum memory is an application of quantum systems, which can preserve quantum information for an extended period of time. Unlike the other quantum computation paradigms and quantum simulation schemes, quantum memory does not aim to perform computation. Therefore, the quantum memory application has different demands on the AQM~\cite{liu2023quantum}.

To support quantum memory applications, \textbf{Rule 1 and 4} of the AQM are required. The AQM \textbf{Rule 2} can be relaxed, as a quantum system does not need to support universal control of individual qubits and a universal gate set for quantum computation. \textbf{Rule 5} can also be relaxed as when the quantum memory acts as a computing module in the future QPU architecture design, its readout capability is not necessary. \textbf{Rule 3} is differently demanded, as the connectivity of the qubits inside the quantum memory modules is not essential, while connectivity to computing qubits is necessary~\cite{liu2023quantum}.

\subsection{Physical systems supporting the AQM} \label{subsec:supporting_AQM}

In this section, we examine the physical system from the lower stacks of the AQM. We do not aim to provide a thorough review of useful physical systems for QC and QIP. Instead, we mainly focus on three types of physical systems: (1) physical systems with strong support of the AQM, e.g., superconducting circuits, trapped ions, and neutral atoms, (2) systems with unique strength and weakness in supporting the AQM, e.g., nitrogen-vacancy centers in diamond crystals, and photonic systems, (3) systems with partial support of the AQM, e.g, the quantum memory systems for photons. 

\subsubsection{Physical systems for AQM support} \label{subsec:sc}

Superconducting circuits~\cite{Devoret2004, Girvin2009, Kjaergaard2020, Kockum2019, Blais2021}, trapped ions~\cite{Haffner2008, Monroe2013, Bruzewicz2019, Brown2021}, and neutral atom systems~\cite{Wu2021, Saffman2010, Bluvstein2022, Bluvstein2024} are widely adopted in QC. The superconducting circuit system is known for its fast and reliable gate operations. The trapped ions system is famous for its connectivity and long coherence while the neutral atom system is good for parallel operations and integration. Although there are other physical systems that also support the AQM well, e.g., quantum dot and solid-state spin qubits, in the interest of conciseness, we mainly focus on the superconducting circuit system to discuss the support of the AQM from the physical stack. Detailed discussions of trapped ions and neutral atoms can be found in the Appendix.

\paragraph{a. Qubit states.} The superconducting qubits use the modes of plasma oscillation as quantum states to encode quantum information. Depending on the ratio of Josephson energy ($E_J$) to the charging energy ($E_C$). When $E_J \ll E_C$, the qubit states have definite Cooper pair numbers, while in the opposite case, the qubit states have a more definite superconducting phase.

Both transmon and fluxonium qubits have higher excited states~\cite{Kjaergaard2020, Blais2021, Manucharyan2009, Koch2007}, while the ions and atoms have other electronic levels. In transmon qubits, due to the anharmonicity provided by Josephson junctions, the transitions to higher excited states can be energetically distinguished from the transition between ground and excited states~\cite{Koch2007, Manucharyan2009}. Although it is still likely to populate higher excited states, especially for transmon qubits, the leakage can be reduced using quantum control methods, e.g., DRAG pulses~\cite{Chow2010}.

In the trapped ion and neutral atoms systems, qubits are made of individual ions and atoms that are trapped in electromagnetic/optical traps. The qubit states can be the electronic levels~\cite{Bermudez2017, Pogorelov2021, Levine2018, Madjarov2020, Picken2019, Morgado2021} or the hyperfine levels of the ion/atom~\cite{Harty2014, Gaebler2016, Ballance2016, Wang2017, Wang2021, Sheng2018, Graham2019, Levine2019, Bluvstein2022}. Ions and atoms also have other levels, which can be distinguished by the energy difference and by the transition rules.

The Hilbert space spanned by the qubit states of transmon, fluxonium, ions, and neutral atoms can be completely addressed using microwave and optical control methods, which will be discussed below in more detail.

\paragraph{b. Operations.} The universal control of a single superconducting qubit can be obtained through external microwave drives~\cite{Chow2010, Koch2007, Manucharyan2009, Houck2008, Blais2021}. Specifically, a superconducting qubit can be driven by microwave fields to apply arbitrary angle Pauli-X and Pauli-Y gates. The Pauli-Z rotations can be implemented virtually. The universal control over a single-qubit Hilbert space can be achieved. 

Two-qubit gates between superconducting qubits can also be implemented. For example, control-phase gates can be implemented by tuning the frequency of one transmon through flux drives~\cite{DiCarlo2010, Barends2013}, while cross-resonance gates between two fixed-frequency transmons or fluxoniums can be implemented using microwave drives~\cite{Rigetti2010, Chow2011, Nesterov2022, Dogan2023}. Furthermore, the recent development of tunable couplers enables tunable couplings between transmon qubits, which enables CZ gates with fidelity reaching $99.8\%$ with about $40$~ns~\cite{Stehlik2021, Sung2021, Marxer2023}. The high-fidelity two-qubit gates can also be realized. An iSWAP gate with gate time $50$~ns can reach fidelity $99.72\%$~\cite{Bao2022}, while microwave-activated CZ gates can be expected to be realized with fidelity $99.9\%$ within $100$~ns~\cite{Nesterov2018, Nguyen2022}. With all these single-qubit and two-qubit gates, a universal gate set is supported.

The universal control of the qubits made of ions or atoms can also be realized using microwave and optical methods~\cite{Hilder2022, Ospelkaus2011, Harty2014, Yavuz2006, Xia2015, Bluvstein2022, Levine2022}. The complete Hilbert space spanned by the qubit states is addressable. Fast and high-fidelity two-qubit gate operations are achieved~\cite{Gaebler2016, Ballance2016, Saner2023, Levine2019, Bluvstein2022, Evered2023, Bluvstein2023, Bluvstein2024}.

Although the gate operations still suffer from errors, there are active explorations on building quantum error correction codes to mitigate the imperfection~\cite{Krinner2022, Google2023, Bermudez2017, Bluvstein2024, Stricker2020}.

\paragraph{c. Connectivity.} However, the coupling between superconducting qubits is limited, where a single superconducting qubit can only couple to the qubits nearby (mostly physically connected). Therefore, the connectivity is usually limited to the nearest neighbors~\cite{Blais2021}. 

Compared to superconducting qubits, the trapped ions and neutral atoms can maintain higher connectivity. Due to the nature of the mechanics of two-qubit gates between ions, it is possible to perform long-range two-qubit gates between ions trapped in a single trap~\cite{MSGate}. In the neutral atom system, the coherent transport of atoms enables shuttling atoms to another site to perform gate operations~\cite{Bluvstein2022, Bluvstein2024}.

\paragraph{d. Long coherence time.} Transmon and fluxonium qubits can preserve a relatively long coherence time compared to the gate time. With the current improvement of material choices and fabrication techniques, the lifetime of transmon qubits has been improved from $\sim 1~\mu$s~\cite{Schreier2008, Houck2008} to $100$~\cite{Barends2013, Chang2013, Jin2015} to $500~\mu$s~\cite{Place2021, Wang2022Transmon}, while the coherence time $T_2^*$ can reach $0.3$~ms~\cite{Place2021, Wang2022Transmon}. The coherence time can be further improved using dynamical decoupling to $0.557$~ms~\cite{Wang2022Transmon}. the fluxonium qubit is less sensitive to charge noise, which extends its coherence time $100$ to $300~\mu$s. Recently, a fluxonium qubit with coherence time $T_2^*$ reaches $1.48$~ms has been reported~\cite{Somoroff2023}

Compared to superconducting qubit systems, trapped ions and neutral atoms can have longer coherence time, especially when the qubit is encoded into the hyperfine levels of the ions/atoms. For example, the coherence time of an ion qubit can reach $5500$~s~\cite{Wang2021}.

\paragraph{e. Qubit readout.} The state of transmon or fluxonium qubits can be read out by dispersive coupling to a microwave field~\cite{Wallraff2004}. The phase accumulated by the microwave field when the qubit is in its ground or excited state is different. Using the phase information of the microwave field, quantum nondemolition measurement on qubit state can be achieved in 40~ns to 100~ns with fidelity 99.0\% to 99.7\%~\cite{Walter2017, Sunada2022, swiadek2023enhancing, Sunada2024}. Although the measurements via this method are only projective measurements along the computational basis, more general measurements, e.g, the stabilizer measurements and error corrections used in quantum error correction codes, can be realized using projective measurements on an auxiliary qubit, which is widely used in quantum error correction~\cite{Krinner2022, Acharya2023}.

Although superconducting qubits fully satisfy the requirements of the relaxed AQM, they still suffer from finite gate errors. In order to reduce the gate errors, several attempts of using superconducting qubits to build error correction codes have been demonstrated~\cite{Andersen2020, Krinner2022, Google2023, Gong2021}. We conclude that superconducting qubits meet all the requirements from the relaxed AQM, and have nearly all mitigation techniques available, which makes superconducting qubits one of the most promising systems for QC. But as the current construction of superconducting qubit QC systems still suffers from finite gate errors and the connectivity in a large integration of qubits, there are opportunities to use superconducting qubits as a small but fast quantum arithmetic unit in future QC design~\cite{liu2023quantum}.

\subsubsection{Nitrogen-Vacancy centers} \label{subsec:nv}

Nitrogen-vacancy centers~\cite{Wrachtrup_2006, Doherty2011, Doherty2013} combined with other solid-state defect centers have emerged as promising candidates for QIP. Defect color centers inside the solid state systems can be nicely fabricated and implanted inside the solid crystal, while they can have long coherence spin states that can be manipulated using microwave and optically readout.

\paragraph{a. Qubit states.}
The negatively charged NV centers have six electrons localized around the defect. The electronic ground state manifold consists of three spin-1 states. When there is no magnetic field applied, the state with $S_1 = 0$ and two states with $S_z = \pm 1$ splits by $2.87$~GHz~\cite{Doherty2011, Doherty2013}. When the NV center electronic spin states are utilized to encode quantum information, a magnetic field is applied to further break the degeneracy of the two states with $S_z = \pm 1$, and use one of them with the $S_z = 0$ state to form a qubit space~\cite{Fuchs2009, Fuchs2010, Abobeih2018}. Another way to encode quantum information is to encode into the nuclear spin states of the nearby atoms (Carbon or Nitrogen atoms)~\cite{Maurer2012, Bradley2019, Pompili2021, Bartling2022, Hermans2022, Xie2023}.

\paragraph{b. Operations.} 
The universal control of a qubit encoded into NV center electronic states can be achieved by coherent microwave drives~\cite{Fuchs2009, Fuchs2010}. The universal control on the nuclear spin states can be achieved by addressing the nuclear spin states by manipulating the electronic states of the NV center via the hyperfine interaction~\cite{Jelezko2004, Abobeih2022}.

However, implementing entangling gates between NV centers to support a full universal gate set for QC is challenging. Entanglement between two electronic spin qubits of NV centers can be entangled by photon heralded entanglement generation process~\cite{Bernien2013, Pfaff2014, Hensen2015, Pompili2021}. Entanglement gates between the electronic spin qubit and the nearby nuclear spin qubits can be performed by microwave drive on electronic spin degree of freedom~\cite{Cramer2016, Taminiau2014, Taminiau2012, Dong_2020, Abobeih2022}.  

With the heralded entanglement between the electronic states of two NV centers, the entangled state can be used to enable gate operations between the nearby nuclear spin states. It can potentially be used to obtain a remote entangling gate for nuclear spin qubits, which can be used to build a quantum network~\cite{Pompili2021}.

\paragraph{c. Connectivity.}

Despite the possibility of heralded entanglement generation, two-qubit entanglement occurs relatively slowly, with a reported rate of $9$~Hz for remote entanglement between two NV centers~\cite{Pompili2021}. Although improvements in photon emission, collection, and detection efficiency could enhance generation speed, the speed of generating entanglement of NV qubits remains slower compared to trapped ion and superconducting systems. 

\paragraph{d. Long coherence time.}
The electronic qubit of NV centers can preserve a relatively long lifetime. Even at room temperature, the NV electronic states can preserve $1.8$~ms coherence time in an isotopically pure diamond crystal~\cite{Balasubramanian2009}. The nuclear spin states can have even longer lifetime~\cite{Maurer2012, Bradley2019, Pompili2021, Bartling2022, Hermans2022, Xie2023}. One challenge of NV center electronic states and nuclear spin states is the relatively short coherence time, which is due to the spin bath of the surrounding nuclear spins in the diamond crystal. However, the effect of the spin bath can be mitigated by dynamical decoupling~\cite{Maurer2012, Bar-Gill2013, Abobeih2018, Abobeih2022, Bradley2019}. With the help of dynamical decoupling, the coherence time of the electronic spin state can reach $1.58$~s~\cite{Abobeih2018}, while the nuclear spin states can be extended to $63$~s~\cite{Bradley2019, Abobeih2022}.

\paragraph{e. Qubit Readout.}
The measurement of NV center electronic states is implemented by optically pumping the states to the excited state manifold, and detecting the emitted photons. Due to the existence of a non-radiative relaxation path from the excited spin-$0$ states, the state of the NV electronic qubit can be distinguished by the photon counts~\cite{Doherty2011, Doherty2013}. As the nuclear spin states are well isolated from the environment, nuclear spin states can be read out by swapping the nuclear state to NV electronic states. Then the state can be detected using the above method~\cite{Abobeih2022}.

Overall, the NV center systems have well-defined and controllable qubit states (\textbf{Rule 1}), long coherence time (\textbf{Rule4}), and easy single-qubit measurement capabilities (\textbf{Rule 5}). However, the connectivity and two-NV entangling gates (\textbf{Rule 2 and 3}) are relatively weakly supported. 

\subsubsection{Photonic qubits for quantum computing} \label{subsec:photon}

Optical photons are widely used in quantum communication and computing. However, their weak nonlinearity in optical nonlinear materials limits fast gate operations compared to microwave photons~\cite{Byer1974, Chen1986nl}, which benefit from Josephson junctions. Another distinct feature of optical photons used in quantum communication and quantum computing is that they are considered `flying qubits', i.e., itinerant photons rather than stationary photons stored in an optical cavity mode. To avoid repeating the discussion covered in the previous sub-sections, we focus on itinerant photons used as qubits in quantum communication and quantum computing, especially in the MBQC model~\cite{Raussendorf2001, Raussendorf2003, Browne2005, Briegel2009}, offering an alternative approach to supporting the AQM.

\paragraph{a. Qubit states.}

There are multiple ways of encoding quantum information into optical photons. For example, the quantum information can be encoded into the polarization of photons (polarization encoding)~\cite{Turchette1995, Knill2001, Duan2005, Brien2003, Basset2021}, the presence or absence of a photon (Fock encoding)~\cite{Chuang1995}, the presence of a photon in early or late time bins (time bin)~\cite{Brendel1999, Jayakumar2014, Sun2022TimeBin, Lee2019, Ortu2022}, the presence of a photon in two modes with different frequencies (frequency encoding)~\cite{Olislager2010, Lu2019, Sabattoli2022, Clementi2023}, etc.

\paragraph{b. Operations.} 
Universal control of a single qubit state varies based on encoding methods. Polarization-encoded photonic qubits can be controlled using passive optical elements like wave plates, beam splitters, phase shifters, and polarizers~\cite{Kok2007}. However, other encoding methods may require active elements and operations. For example, time-bin encoded qubits require delay lines and active optical switches to separate spatial modes, coherently convert photons between modes, and recombine them into time bins~\cite{Kok2007}.

Supporting a universal gate set in photonic systems involves two approaches. One method is implementing an entangling gate, akin to other matter qubit systems. This gate can be executed probabilistically using photon measurements inducing nonlinearity~\cite{Knill2001, Brien2003, Ralph2001, Zou2002, Pittman2002, Pittman2003}, or deterministically utilizing nonlinearity from strongly-coupled cavity-QED systems~\cite{Duan2005, Kimble2008, Zhan2020}. However, quantum gates suffer from imperfections. Probabilistic photon gates can achieve high fidelity through heralding but require more resources to implement~\cite{Kok2007, Brien2003}. Photonic gates using cavity-QED systems also face challenges due to imprecise control and photon absorption, making gate-based photonic QC difficult at the current stage.

On the other hand, universal computing in photonic systems can be implemented by MBQC model~\cite{Raussendorf2001, Raussendorf2003, Browne2005, Briegel2009}. In this model, universal computation is driven by measurements of selected qubits in a certain order on an existing entangled state. The entangled photonic states can be generated by time-delayed feedback~\cite{Pichler2017, Sun2016, Zhan2020, Shi2021}, or fusing small pieces of entangled states together~\cite{Bartolucci2023}, or by photon emission from entangled photonic emitters~\cite{Lindner2009, Economou2010, Buterakos2017, Russo2019}.

The physical support of a universal gate set can be disparate from other matter qubit systems discussed in the rest section. However, a universal gate used in the gate model, e.g., a CNOT gate, can be mapped to a sequence of measurements on a resource state~\cite{Raussendorf2001, Raussendorf2003}. Therefore, unlike gate-based models, after a gate operation, the quantum information is not carried by the same set of physical photonic qubits before the gate. 

\paragraph{c. Connectivity.}
The physical connectivity of optical photons can be easily achieved by guiding the photons together. However, the computation connectivity of photonic qubits is restricted by the capability of performing entangling gates, or the entanglement structure of the resource states for MBQC.

\paragraph{d. Long coherence time.}
As photons do not interact with each other naturally, the photonic qubits can have long coherence times when they propagate in the vacuum. However, when they propagate in optical media, the material absorption causes photon loss, which is the leading error in photonic-based quantum systems. Another source of decoherence is the fluctuations of the optical paths, inducing phase noises to the photonic qubits. Stabilizing optical paths is essential in optical-based quantum systems~\cite{Bernien2013, Pfaff2014, Hensen2015}.

To mitigate photon loss, dual-rail-type encoding methods, such as using two time, frequency, or spatial modes, can enable error detection. Photon detectors can flag errors when no detection event occurs during qubit measurement.

Another way to mitigate the error is to enable error correction on photonic-based QC systems, which have been theoretically proposed and investigated~\cite{Varnava2006, Varnava2007, Buterakos2017, Zhan2020, Hilaire2021, Zhan2023performanceanalysis, Bartolucci2023}. Generating tree graph states for photon loss error has been demonstrated in experiments~\cite{Zhan2020}. However, due to the resource demand, it is still challenging to demonstrate the error correction functionality in photonic systems~\cite{Li2015, Bartolucci2023}. 

\paragraph{e. Qubit readout.}
Photon detection involves capturing photons by detectors, converting them into electric signals (e.g., current)~\cite{scully1997quantum}. Utilizing photon detectors and linear optical devices like wave plates and polarizers, quantum information encoded in photonic qubits can be measured. For example, polarization-encoded photon qubits can be measured using polarizers followed by photon detectors, while the time-bin encoded photon qubits can be detected using optical switches to split them into two spatial modes, and then measured separately. 

A notable distinction between itinerant photonic qubits and matter qubits is that measurement destroys photonic qubits. Unlike matter qubits, where the qubit itself remains alive even after destructive measurement, photonic qubits are absorbed by detectors. This unique feature affects the support for the AQM by photonic systems. Resetting photonic qubits through measurements is not physically feasible, but they can be replenished once the measurement result is known. 
In quantum algorithms using photonic systems, the number of photons qubits can exceed the number of logical qubits, especially when there are mid-circuit measurements in the algorithm. Multiple photonic qubits can represent a single computational qubit in certain quantum algorithms. 

To summarize, although the physical photons may not remain alive during the computation, as the photons can potentially be replenished, \textbf{Rule 1} of the AQM can still be supported. As the photonic two-qubit gates are still challenging to perform, \textbf{Rule 2} of the AQM can only be weakly supported. \textbf{Rule 3} of the AQM is hard to be supported by photonic qubits in terms of gate operations, however, can be supported with the aid of photon emitters. \textbf{Rule 4 and 5} of the AQM are well supported by photonic systems.

\subsubsection{Quantum memory systems for photons} \label{subsec:qm}

At the end of this section, we consider a type of physical system well-suited for buffering optical light, which is used as optical quantum memories. These systems cannot support the AQM, however, still useful for QIP applications.

\paragraph{a. Qubit states.}

There are different constructions of optical memories. Quantum memories using single atoms, or defect centers in solid-state systems can also be utilized as a computational qubit, as we discussed in Sec.~\ref{subsec:nv}. In this section, we examine two main types: atomic clouds and rare-earth-ion-doped crystals. Both types store quantum information in collective excitations of atoms or ions, rather than individual ones~\cite{Lvovsky2009, Heshami2016, Fleischhauer2005, Lukin2003, Ledingham2012, Ferguson2016, Alexander2006, Gorshkov2007}.

The qubit states in these systems can be defined by the presence or absence of such collective excitations. However, universal control over the Hilbert space encompassing these states is challenging. Therefore, while these systems offer two distinguishable states for quantum information storage, they may not support two states for the AQM.

\paragraph{b. Operations.} 

Realizing universal control of the Hilbert space spanned by the two states requires the collective control of all the atoms/ions inside the systems, which hinders the arbitrary transformations in the Hilbert space. However, by applying external optical light or electronic voltage to control the absorption of the atoms/ions, the incoming photons can be absorbed and converted into the collective excitation of the material, and then the absorbed photon can be re-emitted. Several techniques have been developed for these two operations, including electromagnetically-induced transparency (EIT)~\cite{Fleischhauer2005, Lukin2003, Hosseini2011, Dudin2013, Ding2013, Nicolas2014, Ding2015, Ding2015Nat, Parigi2015, Saunders2016, Katz2018, Hsiao2018, Jiang2019, Wang2019, Li2020, Dideriksen2021, Wang2022, Messner2023, Buser2022}, controlled reversible inhomogeneous broadening (CRIB)~\cite{Moiseev2001, Nilsson2005, Kraus2006, Alexander2006, Hetet2008}, atomic frequency combs (AFC)~\cite{McAuslan2012, Saglamyurek2011, Clausen2011, Ledingham2012, Zhou2012, Ferguson2016, Jin2022, Ma2021_v2, Seri2017, Kutluer2017, Laplane2017, Holzapfel2020, Businger2020, Askarani2021, Ma2021}, and rephased amplified spontaneous emission (RASE)~\cite{Ledingham2012, Ledingham2010}.

\paragraph{c. Connectivity.}

The multi-mode feature of the collective excitation used in atomic clouds and doped ions enables the integration of multiple memory cells into a single physical system~\cite{Vernaz-Gris2018, Messner2023}. However, due to the control complexity, entangling different modes can be hardly implemented. Therefore, the connectivity of different memory modes is limited.

In contrast, two memory systems can be physically connected through optical paths, where the optical light stored in one memory can be emitted and absorbed by the connected memory system. Routing the optical photons can be achieved by active optical elements, which can potentially configure the memory system connectivity.

\paragraph{d. Long coherence time.}

The collective quantum state of the ensemble of atoms and ions can preserve the quantum information for a long time. For example, the $1/e$ decay lifetime of the stored light can reach $16$~s with dynamical decoupling in a cold atomic cloud using the EIT technique~\cite{Dudin2013}, while an AFC-based ion-doped crystal system an obtain storage lifetime $52.9$~min with dynamical decoupling~\cite{Ma2021}.

\paragraph{e. Qubit readout.}

Direct measurement of the collective quantum states is intimidating to realize due to the difficulty of controlling the quantum states of all the atoms/ions simultaneously. However, the measurement can be possible by detecting the state of the emitted photons. 

Overall, the quantum memory systems have good support of \textbf{Rule 1 and 4} of the AQM, as they have good coherence time while they have quantum states that encode quantum information. However, \textbf{Rule 2} is not supported, as universal gate sets are not available. The AQM \textbf{Rule 3} and \textbf{Rule 5} are partially supported, as the direct entangling operations and measurements on these quantum memory systems are challenging.

\subsection{Physical systems with co-design opportunities} \label{subsec:codesign}

The AQM facilitates efficient application-device co-design by aligning quantum algorithm demand with physical system support. When the requirements of quantum algorithms and the capabilities of the physical system supporting the AQM coincide, the application can be performed on this physical system. We illustrate a few examples.

As the photonic system is challenging to perform two-qubit gates between physical qubits, \textbf{Rule 2 and 3} of the AQM is not well supported between physical qubits. According to the unique support of the AQM by photonic qubit systems, MBQC is well-suited~\cite{Bartolucci2023, Kok2007}. In addition, although the entangling gates between photons are challenging, the photon hopping between coupled modes can be relatively easy to achieve, which provides the necessary connectivity (\textbf{Rule 3}) requirements for the application, such as quantum random walk models~\cite {Tang2018} and sampling problems~\cite{Zhong2020, Deng2023}.

In contrast, NV center systems have long coherence times (\textbf{Rule 4}) and good support of the AQM \textbf{Rule 1 and 5}. Although the support of \textbf{Rule 2 and 3} are limited, they are promising candidates for quantum memory and communication applications where a universal gate set isn't necessary~\cite{Bernien2013, Pfaff2014, Hensen2015, Borregaard2020, Pompili2021, Jing2022, Ruf2021, liu2023quantum}. Similar to the quantum memory systems for photons, where \textbf{Rule 2 and 3} are not supported. In addition, the quantum memory systems for photons also have limited support to the readout capability (\textbf{Rule 5}). However, referring to Table.~\ref{tab:algorithms}, they can be well suited to be used as quantum memory, which can not only be used in quantum communication and quantum networks~\cite{Askarani2021, Jobez2016}, but also in quantum memory of a heterogeneous QC architecture~\cite{liu2023quantum}. 

Superconducting, trapped ions, and neutral atom systems offer support for the complete AQM, making them promising candidates for not only universal quantum computing but also applications with stronger partial AQM demands on specific aspects. Compared to the superconducting qubit systems, trapped ions and neutral atoms have higher connectivity and longer coherence time, which also enable them to be used as quantum interconnect modules or quantum memory modules in future heterogenous QC systems with high fidelity coupling~\cite{stein2023microarchitectures, liu2023quantum}. In addition, the high connectivity also makes them suitable for quantum annealing and quantum simulation applications~\cite{Daley2022, Monroe2021, Ebadi2021}.

\section{Conclusion} \label{sec:summary}

In conclusion, we refine the abstract qubit model, addressing the demands posed by quantum algorithms and applications, as well as the requirements of physical systems. However, achieving the ideal AQM is challenging for state-of-the-art physical systems. Instead, these systems can meet a less stringent set of requirements, utilizing techniques such as unitary decomposition, quantum error mitigation, and error correction methods, to compensate for the partial support of the AQM, albeit at the cost of more time and resources.

Furthermore, we observe that physical systems can offer unique advantages in supporting the AQM and quantum applications, especially for quantum algorithms and applications that don't need the complete AQM. This presents co-design opportunities to utilize the strengths of current physical systems by breaking the AQM abstraction layer. We discuss examples such as quantum annealing, quantum random walk, measurement-based quantum computing, analogue quantum simulation, and quantum memory applications, focusing on their specific demands on the AQM. Subsequently, we discuss how physical systems like superconducting qubit systems, NV centers, photonics, and quantum memory materials can support the AQM and their potential roles in future QC system designs. 

As we have shown, the AQM can be a flexible model, which can be adjusted to better support various quantum applications. Therefore, we believe that the AQM can be a useful tool for future quantum algorithm-device co-design. The AQM discussion can guide QC and QIP researchers in developing new algorithms and physical devices, and take benefits of quantum co-design to enhance the capabilities of existing QC devices.

\section{Acknowledgement}


This work was solely supported by U.S. Department of Energy, Office of Science, National Quantum Information Science Research Centers, Codesign Center for Quantum Advantage (C2QA) under contract number DE-SC0012704, (Basic Energy Sciences, PNNL FWP 76274)

\bibliography{ref}
\bibliographystyle{quantum}

\onecolumn
\appendix

\section{Supporting the AQM by trapped ions}
Trapped ion systems support well on the AQM. For a thorough review of trapped ion systems for QC and QIP, we suggest referring to Refs.~\cite{Haffner2008, Monroe2013, Bruzewicz2019, Brown2021}. 

\paragraph{a. Qubit states.}
In a trapped ion system, each ion inside the trap can be used as a qubit. The quantum information can either be encoded into the hyperfine levels~\cite{Harty2014, Gaebler2016, Ballance2016, Wang2017, Wang2021} or Zeeman sublevels of a same orbital~\cite{Keselman2011, Ruster2016, Hilder2022}, or other quantum states in the specific ion level structures~\cite{Bermudez2017, Toyoda2010, Pogorelov2021}. 

\paragraph{b. Universal control.} 
Depending on the type of ion qubits, quantum manipulation schemes for qubit universal control of the ion qubits also vary. Specifically, for hyperfine ion qubits, single-qubit gates can be implemented using either optical Raman transitions~\cite{Hilder2022} or microwave drives~\cite{Ospelkaus2011, Harty2014}. The microwave Pauli-X and Pauli-Y rotations with fidelity $\sim 99.9999\%$ have been reported~\cite{Harty2014}. The microwave gates duration has been improved to $\sim 1~\mu$s without sacrificing much gate fidelity~\cite{Leu2023}. Raman transition-based single-qubit gates achieve fidelity $99.993\%$ in $7.5~\mu$s~\cite{Ballance2016}. Using ultra-fast laser pulses to strongly drive Raman transition can achieve a $\pi$-pulse in $\sim 50$~ps with fidelity $\sim 99\%$~\cite{Campbell2010}. These gates provide the necessary tools for universal control of single qubits.

The coupling between two ions can be induced using the Coulomb interactions between them. There are several schemes to perform two-qubit gates between trapped ions, e.g., the Cirac-Zoller gate~\cite{CiracZoller}, the M\o{}lmer-S\o{}rensen (MS) gate~\cite{MSGate}, and the Leibfried geometric phase gate~\cite{Leibfried2003}. Specifically, MS gate utilizes the phonon modes of the trapped ion chain to mediate the coupling between different trapped ions, which is widely used in trapped ion systems~\cite{Gaebler2016, Blumer2021, Moses2023, Jeon2023, Saner2023}, while the fidelity of two-qubit gates can reach $99.9\%$~\cite{Ballance2016}. The two-qubit gates can be performed in $10$ to $500~\mu$s~\cite{Gaebler2016, Ballance2016, Saner2023}. The two-qubit entangling gates combined with the single-qubit universal control provide universal gate sets for QC.

\paragraph{c. Long coherence time.}
Depending on the type of encoding, the coherence time of the qubits can be different. For instance, in the Zeeman qubits, the coherence time can reach $300$~ms~\cite{Ruster2016}, while hyperfine states are more coherent, and hence the coherence time of a hyperfine qubit can reach several minutes to an hour~\cite{Harty2014, Wang2017, Wang2021} ($5500$~s reported in Ref.~\cite{Wang2021}).

\paragraph{d. Qubit measurements.}
The qubit measurement is achieved using optical approaches, where one of the ion qubit states is excited to an optical active state, such that the relaxation of the ion can emit optical photons for detection. Therefore, by distinguishing the collected emitted photons, the state of the ion can be determined. This physical process is a projective measurement of the qubit computation states. The measurements on ion states can be obtained with fidelity $> 99.9\%$~\cite{Burrell2010, Christensen2020, Todaro2021}.

\paragraph{e. Connectivity.}
The ions are trapped using radio-frequency Paul traps~\cite{Neuhauser1980} and other types of electromagnetic traps~\cite{Dehmelt1990, Chiaverini2005, Seidelin2006, Labaziewicz2008}. The trapped ion systems can have all-to-all connectivity between the ions trapped in a single trap. This is due to the mechanism of two-qubit gate operations. The M\o{}lmer-S\o{}rensen gate requires using the collective motion (phonon) modes to mediate the coupling, which enables this all-to-all connectivity. However, limited by the size of the ion trap and the distinguishability between different phonon modes, it is impossible to have millions of ions trapped in a single trap and have all-to-all connectivity. Furthermore, due to the 1D nature of the ion traps, it is hard to maintain a higher-dimensional ion array to improve integration and connectivity.

With the techniques of ion shuttling, it is possible to select the ion qubits that are necessary to be coupled in the quantum algorithm, and shuttling the ion qubits into the same ion trap to enable the coupling~\cite{Moses2023}. This technique potentially increases the connectivity of the ion qubits. However, shuttling operations can increase the decoherence noise and time consumption.

In summary, trapped ion systems also support the AQM well. Similar to the superconducting qubit systems, the imperfection of the gate operations and decoherence limit the support to an ideal AQM. With the quantum error correction codes, it is possible to overcome the errors in the quantum operations. Compared with superconducting qubits, trapped ions can have longer coherence time, and all-to-all connectivity between ions in the same trap, which enables potential application to quantum memory and quantum simulation~\cite{Zhang2017, Lv2018, Monroe2021, Kiesenhofer2023}.

\section{Supporting the AQM by neutral atoms} \label{subsec:rydberg}

Trapped neutral atom systems are becoming increasingly popular with the new progress of showing the potential of enabling error correction functionalities. Neutral atoms can also have good support for the AQM.

\paragraph{a. Qubit states.}
Neutral atom systems use the spin-electronic states of the trapped atoms to encode quantum information, similar to the ion qubits in the trapped ion system. There are a few strategies to encode quantum information into states of Rydberg atoms. For instance, the ground state of the atom and its Rydberg excited state can be used to encode the qubit state $\ket{0}$ and $\ket{1}$ states~\cite{Levine2018, Madjarov2020}, which is referred to as GR qubits in Ref.~\cite{Picken2019, Morgado2021}, while two hyperfine ground states of the Rydberg atoms can also be used~\cite{Sheng2018, Graham2019, Levine2019, Bluvstein2022}, which is called GG qubits in Ref.~\cite{Morgado2021}). The Hilbert space spanned by the states in different selections can also be fully addressed.

\paragraph{b. Universal control.} 
The single-qubit universal control can be achieved using optical Raman transitions or using microwave drives with optically activated Stack shifts~\cite{Yavuz2006, Xia2015, Bluvstein2022, Levine2022}. The Rabi rate of $2$~MHz has been realized for single-qubit gate operations~\cite{Levine2022}. With these control methods, the Hilbert space can be fully addressed.

To support a universal gate set for QC, entangling gates between two qubits are necessary. The Rydberg interactions can be leveraged to perform entangling gates. When the atom is excited to a highly excited state (Rydberg state), the radius of the Rydberg state is much larger than the radius of the atom in the ground state, which activates a strong dipole-dipole interaction between Rydberg atoms. Specifically, the dipole interaction makes an excited Rydberg atom strongly shift the energy level of the other Rydberg atoms, which blocks the excitation of the nearby atoms. This effect is named ``Rydberg blockade''~\cite{Lukin2001, Jaksch2000, Urban2009}, which enables fast two-qubit gates. The two-qubit gates with fidelity $97.4\%$~\cite{Levine2019, Bluvstein2022} and $99.5\%$~\cite{Evered2023, Bluvstein2023} have been demonstrated in experiments. Specifically, the CZ gate between two atoms can be implemented in $\sim 200$~ns, which greatly suppresses the error from the decoherence of the Rydberg state~\cite{Evered2023}.

\paragraph{c. Long coherence time.}
The coherence time of the qubits varies according to the species of qubits and the trapped atoms, ranging from a few microseconds to a few seconds~\cite{Norcia2019, Ebadi2021, Jenkins2022, Ma2022}. The coherence time can be further enhanced using dynamical decoupling to $1.5$~s for Rb atoms~\cite{Bluvstein2022, Bluvstein2023} and $3.7$~s for Yb atoms~\cite{Jenkins2022}.

\paragraph{d. Qubit measurements.}
The measurement of the trapped neutral atoms is similar to the method used in trapped ion systems. Fast measurements with descent measurement fidelity on the qubit state have been demonstrated, which can be used in quantum error correction~\cite{Xu2021Rydberg, Singh2023, Deist2022, Graham2023}. 

\paragraph{e. Connectivity.}
Compared to trapped ion systems, the neutral atoms cannot be trapped using RF electromagnetic traps. Instead, neutral atoms are optically trapped into optical lattices. With the optical trapping techniques, higher dimensional optical lattices have been realized~\cite{YWang2015, YWang2016, Xia2015, Maller2015, Graham2019}. In addition, due to the large radius of the atom Rydberg state (can reach a few micrometers~\cite{Urban2009, Gaetan2009}), the entangling gates can be applied to two nonadjacent atoms, which potentially increase the connectivity of the neutral atom qubits.

Furthermore, atoms can be precisely transported with high fidelity by adjusting the optical traps~\cite{Ebadi2021, Bluvstein2024}. Through this shuttling process, atoms trapped in distant locations can be brought to nearby sites, implementing entangling gate operations. This significantly enhances the connectivity of neutral atom systems.

In essence, neutral atom systems show promise as a robust platform for AQM. Like superconducting qubits and trapped ions, a key challenge lies in developing error correction codes and demonstrating error suppression for full AQM support. However, the high connectivity of atoms in neutral atom systems makes them particularly suitable for quantum simulation applications~\cite{Ebadi2021, Wu2021, Daley2022, Gross2017}.

\end{document}